# Environmental Factors Can Have Opposite Biodiversity Influences on the Community Temporal Stability In Aquatic Ecosystems

Zihao Wen[1,#], Hang Shan[1,#], Hao Wang[1], Yu Cao[2], Liang He[3], Wenjing Ren[1], Chengjie Yin[1], Qingchuan Chou[1], Chaochao Lv[1], Haojie Su[4], Tao Tang[1], Qinghua Cai[1], Leyi Ni[1], Wen Xiao[5], Xiaolin Zhang[1], Kuanyi Li[6], Te Cao[1,*], Ming-Chih Chiu[1,*], Vincent H. Resh[7], Pablo Urrutia-Cordero[8]

[1] State Key Laboratory of Freshwater Ecology and Biotechnology, Institute of Hydrobiology, Chinese Academy of Sciences, Wuhan, 430072, China

[2] Key Laboratory of Aquatic Botany and Watershed Ecology, Wuhan Botanical Garden, Chinese Academy of Sciences, Wuhan, 430074, China

[3] School of Resources, Environmental and Chemical Engineering and Key Laboratory of Poyang Lake Environment and Resource Utilization of Ministry of Education, Nanchang University, Nanchang, 30031, China

[4] Institute for Ecological Research and Pollution Control of Plateau Lakes, School of Ecology and Environmental Science, Yunnan University, Kunming, 650500, China

[5] Institute of Eastern-Himalaya Biodiversity Research, Dali University, Dali, 671003, China

[6] State Key Laboratory of Lake Science and Environment, Nanjing Institute of Geography and Limnology, Chinese Academy of Sciences, Nanjing, 210008, China

[7] Department of Environmental Science, Policy & Management, University of California Berkeley, Berkeley, USA

[8] IMDEA Water Institute, Science and Technology Campus of the University of Alcalá, Avenida Punto Com 2, Alcalá de Henares, Madrid, 28805, Spain

**Author Contributions**

Te Cao, Leyi Ni, and Xiaolin Zhang conceived the study. Zihao Wen, Hang Shan, Hao Wang, Wenjing Ren, Chengjie Yin, and Chaochao Lv run field studies. Zihao Wen analysed the results and wrote the manuscript. Yu Cao, Liang He, Qingchuan Chou, Haojie Su, Tao Tang, Qinghua Cai, Wen Xiao, Kuanyi Li, Ming-Chih Chiu, Vincent H. Resh, and Pablo Urrutia-Cordero provided valuable comments. Hang Shan, Ming-Chih Chiu and Pablo Urrutia-Cordero supervised data analyses. All authors revised the manuscript.

**Data Availability Statement**

The authors confirm that the data and the code supporting the results are available online on the figshare platform at: https://figshare.com/account/items/27957372/edit

**Conflicts of Interest**

The authors declare no conflicts of interest.

**Acknowledgments**

This study was supported by the Key Program (Grant No. 31930074), the Normal Project (Grant Nos. 32201340; 32071574) of the National Science Foundation of China, and the State Key Laboratory of Freshwater Ecology and Biotechnology (Grant No. 2019FBZ01). We thank Erik Jeppesen for his valuable comments on the revision of the manuscript.

**Abstract**

1. An understanding of how biodiversity confers ecosystem stability is crucial in managing ecosystems under major environmental changes. Multiple biodiversity drivers can stabilize ecosystem functions over time. However, we know little about how local environmental conditions can influence these biodiversity drivers, and consequently how they indirectly shape the ecological stability of ecosystems.

2. We hypothesized that environmental factors can have opposite influences (i.e., not necessarily either positive or negative) on the temporal stability of communities in different environmental ranges depending on the biodiversity drivers involved. We tested this novel hypothesis by using data from a 4-year-long field study of submerged macrophyte across a water depth gradient in 8 heterogeneous bays of Erhai lake (with total sample size of 30,071 quadrats), a large lentic system in China.

3. Results indicate that a unimodal pattern of stability in temporal biomass measurements occurred along the water-depth gradient, and that multiple biodiversity drivers (the asynchrony in species dynamics, and the stability of dominant species) generally increased the temporal stability of aquatic primary producers. However, the effect of water depth either increased or decreased the stability of biomass according to the environmental conditions associated with sites along the water depth gradient.

4. Synthesis. These results reveal the influence of local environmental conditions on the biodiversity drivers of stability may help predict the functional consequences of biodiversity change across different scenarios of environmental change.

**Introduction**

The interaction of biodiversity and temporal stability, which reflects the degree of ecological variables that remain unchanged over time, has been a subject of interest to ecologists through theoretical, field, and experimental studies (e.g., Hooper *et al.*, 2005; Tilman *et al.*, 2006; Yan *et al.*, 2023). Based on those efforts, biodiversity has been found to promote temporal stability of the community through key mechanisms such as: (1) species asynchrony, where declines in some species are offset by increases in others (Gonzalez & Loreau 2009); (2) dominant species stability, where stable species enhance community performance (Grime 1998; Yan *et al.*, 2023); and (3) species richness, which fosters stability via mechanisms like the portfolio effect (i.e., the statistical averaging of independent fluctuations of individual species at community level), especially in communities with evenly distributed abundances of species (Doak *et al.* 1998; Tilman *et al.*, 1998). Although an understanding of the above biodiversity drivers is crucial to maintaining ecosystem functions and services (Hooper *et al.* 2005), environmental processes can alter the mechanisms by which biodiversity confers stability to ecosystems (Xu *et al.* 2015).

The prevailing focus on the environmental mechanisms that drive ecological stability highlights how environmental conditions may modulate the biodiversity-stability relationship (Garcia-Palacios et al., 2018, Gilbert et al., 2020). For example, although the biodiversity-stability relationship is context-dependent (Hallett et al., 2014, Garcia-Palacios et al., 2018, Sasaki et al., 2024), biodiversity is hypothesized and has been demonstrated to positively affect the temporal stability of the community, regardless of environmental conditions (Hong et al., 2022, Sasaki et al., 2024). Related to this, Garcia-Palacios et al. (2018) reported that species richness can enhance the stability of plant communities under low aridity conditions yet still have a stabilizing influence, even under the most aridity conditions.

As a different, novel perspective, we argue that environmental factors can produce different and even opposite biodiversity-mediated influences (i.e., but not restricted to strictly positive or negative) on the temporal stability of communities. Biodiversity drivers, including species richness and species asynchrony, can exhibit multiple nonlinear patterns both positively and negatively over a full range of environmental gradients in terrestrial, freshwater, and marine ecosystems (e.g., Quintero & Jetz 2018; Dodds *et al.*, 2019; Chaudhary *et al.*, 2021), such as with multimodal or unimodal elevational patterns of biodiversity (Peters *et al.*, 2016). For example, in a low aridity range, aridity positively correlates with the species richness of plants, whereas in a high aridity range, aridity is negatively associated with plant species richness (Garcia-Palacios *et al.*, 2018). Gilbert *et al.* (2020) found that the moderate levels of climate variability can increase the asynchrony of plant species, but that high levels of variability cause a decline in species asynchrony. When considering the positive biodiversity-stability linkages in diverse ecosystems (e.g., Fu *et al.*, 2021; Ouyang *et al.*, 2021; Quan *et al.*, 2021; see also the previous paragraph), opposite relationships are expected to occur between environment and community stability along the environmental gradients (Figure 1a).

In this study, we used monitoring information for submerged macrophytes to examine how environmental conditions may influence the temporal stability of communities via biodiversity drivers across environment gradients in 8 heterogenous bays of a 40-km-long lake in China. We develop a generalized hypothesis that the environment can have opposite effects (i.e., not limited to strictly positive or negative) on community stability in different environment gradient ranges through the biodiversity drivers influencing the ecosystem (Figure 1a). In lake ecosystems, we hypothesize that environmental conditions (i.e., stresses caused by water level fluctuations and restrictions in light availability) will shape the unimodal pattern of biodiversity of submerged plants along a water-depth gradient (Figure 1b), for example, in species richness (Spence 1982; Ye *et al.*, 2018; Lewerentz *et al.*, 2021). Based on this unimodal environment-biodiversity relationship, we also hypothesize that when biodiversity is considered to positively affect stability, the water depth will consequently be positively correlated with community stability in shallow water and negatively correlated in deep water (Figure 1c).

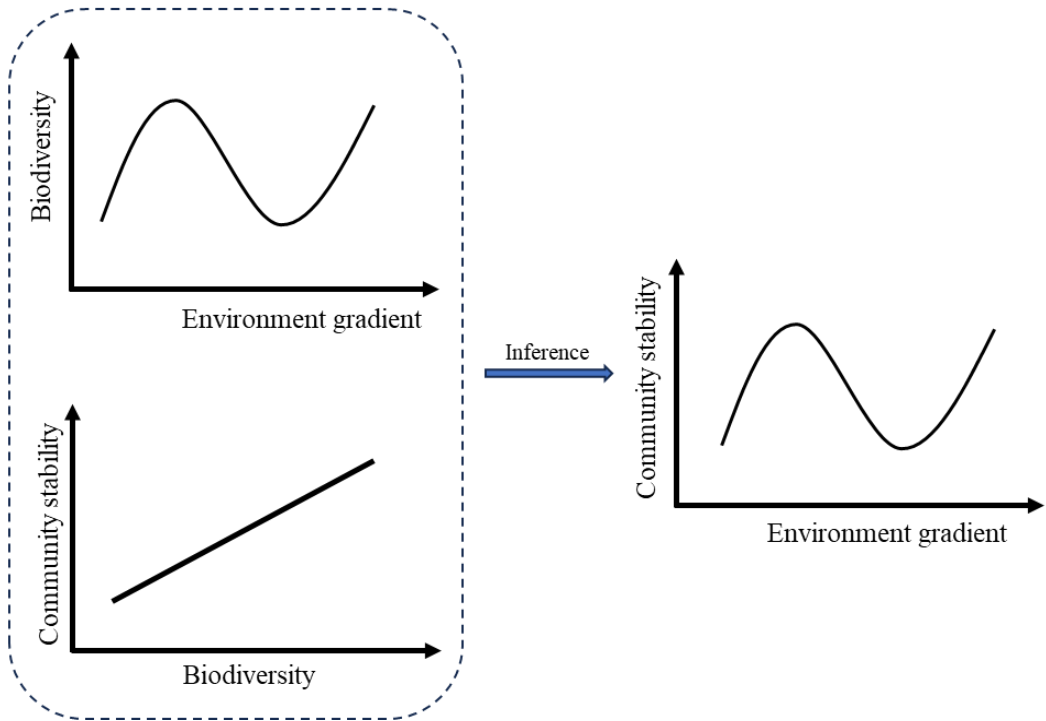
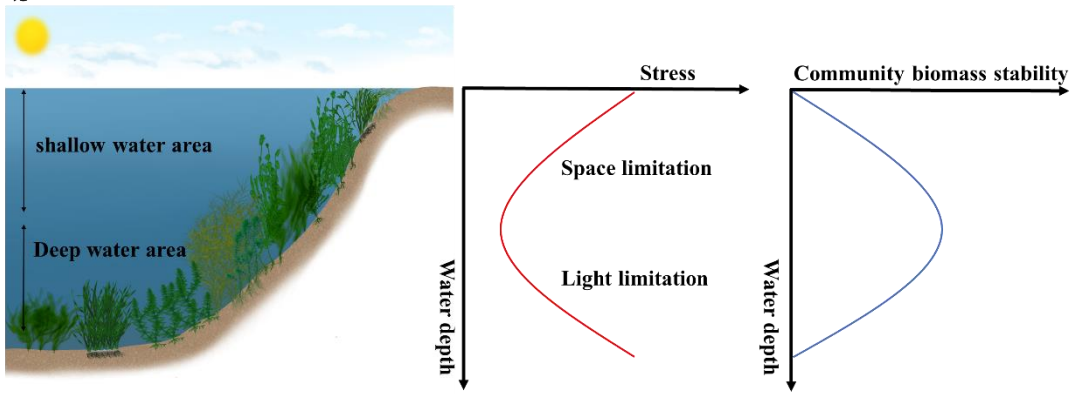

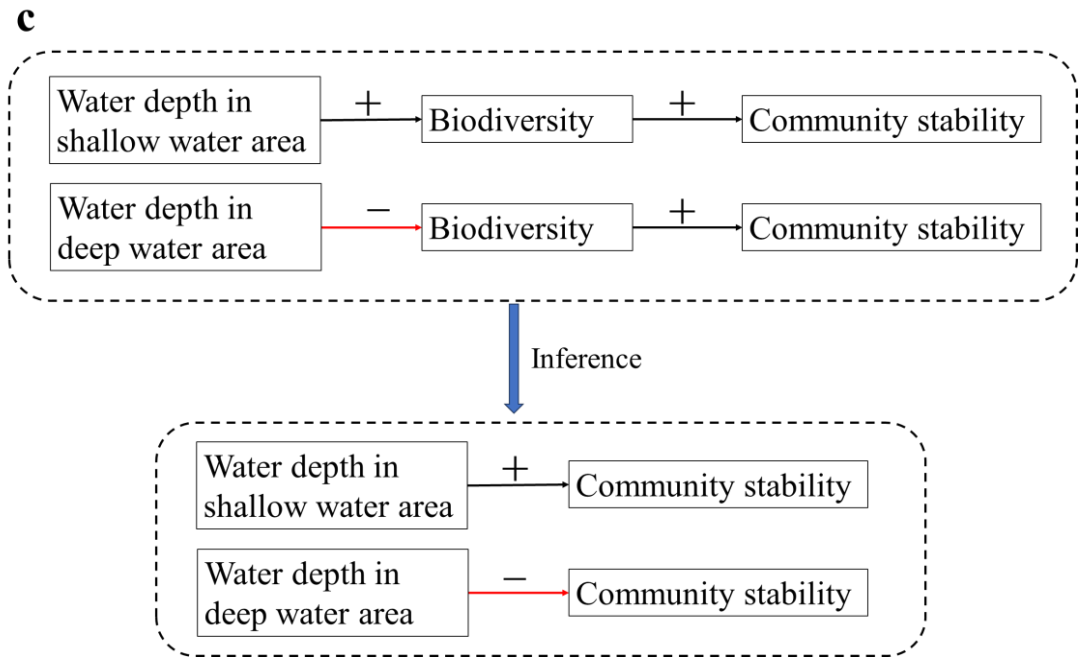

**Figure** 1. (a) Conceptual diagram of the non-linear relationship, which combines positive and negative phases (e.g., multimodal or unimodal), between ecological stability and environment gradients. The environment can have opposite effects (i.e., not necessarily either positive or negative) on temporal stability, depending on the positive and negative influences of environmental factors on the biodiversity drivers along different environmental ranges. This is particularly true when the positive relationship between biodiversity and stability persists under a range of conditions. (b) Conceptual representation of submerged macrophytes experiencing space and light limitations along depth gradients. In shallow water, macrophytes face seasonal exposure and disturbances from water level fluctuations; light availability exceeds the plant's needs. In deep water, limited light restricts their growth and biomass, but there is no seasonal exposure as in shallow water. Consequently, there is an optimum of biomass stability over a range of water depths. (c) In contrast, water depth has the opposite effects on temporal stability depending on the positive or negative influence on the biodiversity drivers of temporal stability across the range of water depths.

**Materials and methods**

*Study Area*

This study was conducted in Erhai Lake (25°36′–25°58′N and 100°15′–100°18′E), located northwest of the Yunnan Plateau in China. With a water surface area of 252 km$^2$, average depth of 10.5 m, and maximum depth of 20.5 m, it is the second-largest lake in Yunnan Province. The area has a subtropical, monsoon climate. The mean annual temperature is 15.6 °C, with a maximum air temperature of 20.1 °C in July and minimum temperature of 8.7 °C in January. The mean annual precipitation is 870 mm, of which > 80% falls between May and October (Wen *et al.*, 2021). Erhai Lake is in the early stages of eutrophication (Wen *et al.*, 2022). Submerged macrophyte species are show zonation distribution along the water depth gradient, with dominant species including *Potamogeton maackianus*, *Ceratophyllum demersum*, and *Vallisneria natans* (Wen *et al.*, 2022).

*Study design and sampling protocol*

We carried out monthly monitoring of eight bays (Shaping Bay, Xizhou Bay, Majiuyi Bay, Erhaiyue Bay, Xiangyang Bay, Wase Bay, Changyu Bay, Hongshan Bay) around Erhai Lake from June 2017 to May 2021 in order to estimate the biomass of individual species and the total community biomass along water depth gradients (Figure Supplementary 1). We used a high spatial resolution survey method done along a water depth gradient in eight heterogenous bays in a 40 km long part of the lake. In the area studied, the two long sides of the lake were influenced by river inputs from different parallel mountain systems. The lake areas were situated more than 10 km apart along each side of the lake and were exposed to varying land-use inputs (more details in Figure Supplementary 1).

A certain range of water depths in this lake allows submerged macrophytes to survive, and the maximum colonization depth in this lake is generally less than 7 m (Middelboe & Markager 1997). The sampling method that we used for submerged macrophytes covers the maximum range of the water depth of submerged plants that can grow in Erhai lake, which is no more than 7 m.

We established three permanent transects in each bay that were perpendicular to the shoreline. For each transect, three 5 × 5 m plots were located along a 0 to 7.0 m depth gradient at 0.5-m intervals. In this study, the 48-month average water level was 1,965.1 m (Supplementary Figure 3). For sampling, lake bottom elevations were divided into 0.5 m intervals from 1,958.1 to 1,965.1 m, corresponding to depths of 6.5–7, 6–6.5, 5.5–6, and so on down to 0–0.5 m.

Within each 25 m² plot, we used three 0.2 m² quadrats for analysis. A submerged rotatable reaping hook covering a bottom area of 0.2 m² was used to collect submerged macrophytes from each quadrat. The collected plants were washed, identified, sorted by species, and weighed (biomass in fresh weight). Plants were classified into 3 groups (dominant, common, and rare species) based on their relative abundance. Relative species abundance is the ratio between the biomass of each species and the biomass at the community level. Dominant species included those with a relative biomass > 10%, common species were those with a relative biomass e ranging from 1% to 10%, and rare species were those with a relative biomass < 1% in the sampling plots over the entire study period (Ma *et al.* 2017). In total, we sampled a total of 30,071 quadrats.

*Ecological stability and biodiversity drivers*

Species richness in each plot was defined as the total number of species detected in each quadrat across 4 years. Temporal stability of the community was defined as μ/σ, where μ and σ are the mean biomass and temporal standard deviation of community biomass at each water depth interval over the four-year period (Hautier *et al.* 2014).

We also calculated Simpson's dominance index by using the mean biomass of each species across 4 years of each plot of each water depth gradient. Simpson's dominance index was calculated as follows:

$$\text{Simpson} = \sum_{i=1}^{n} \left(\frac{b_i}{B}\right)^2 \tag{1}$$

Where i represents the number of species in the plot, 1≤i≤n, $b_i$ is the biomass of species i, B is the ecosystem biomass with a plot of n species (Leps 2004).

Species asynchrony, which represents the degree of different responses of species within a community to environmental change (Loreau & de Mazancourt 2013), was calculated at each water depth interval over the four-year period The degree of species

asynchrony was quantified by the community-wide asynchrony index as follows:

$$1 - \varphi_x = 1 - \sigma^2 / (\sum_{i=1}^{S} \sigma_i)^2 \qquad (2)$$

where $\varphi_x$ is the species synchrony, $\sigma^2$ is the community biomass variance, and $\sigma_i$ is the standard deviation of biomass of species *I* in a plot with *S* species. This index equals 1 when species fluctuations are totally asynchronized and equals 0 when species fluctuations are totally synchronized.

The R package 'vegan' was used to calculate the Simpson's dominance index. The 'synchrony' function in the R package 'codyn' was used to calculate synchrony metrics (Hallett et al. , 2016).

*Statistical analysis*

In order to categorize the two water depths that could affect the potential mechanisms driving community stability, we applied a break-point analysis within a piecewise linear regression analysis. In the analysis, each sample represents a value of community biomass stability at each water depth gradient along the depth gradient. Such break-point analysis estimates the segment with the largest change in the temporal stability of the community biomass along the depth gradient (Muggeo 2003). The community biomass stability of submerged macrophytes was calculated for two water depth areas, the shallow water and deep areas.

To test whether species richness enhances community stability through the mean-variance scaling, we examined the mean-variance scaling relationship (Taylor's power law) on community biomass stability (Doak *et al.*, 1998; Tilman *et al.*, 1998). Increases in species richness can create a "portfolio effect" (also known as "statistical averaging"), where stability at the community level increases as a result of more independent fluctuations in species' abundances (i.e., fluctuations with no correlation) (Doak *et al.*, 1998). This probabilistic (and non-biologically related) process can be described by the mean-variance scaling relationship. The relationship was described as:

$$\sigma^2 = cm^z \qquad (3)$$

where $\sigma^2$ is the variance in biomass per species, c is a constant, m is the average biomass per species, and Z is the scaling coefficient. When $1 < Z < 2$, diversity is expected to enhance the stability of the plant community biomass via mean-variance scaling (Tilman *et al.*, 1998; Grman *et al.*, 2010). A linear regression model was used to explore the relationship between log (species variance of biomass) and log (species mean biomass) for different water depth intervals. ANCOVA analyses were used to examine differences between the slopes of the linear regressions for different water depth areas, indicated by a significant difference in slopes at the 0.05 level

General Additive Models (GAMs) were used to explore separately the relationship between water depth and multiple biodiversity facets (species richness, species asynchrony, species dominance, and the biomass temporal stability of dominant, common, and rare species). In addition, we used GAM models to examine how different species were distributed along the water depth. We used the GAM approach because it does not require *a priori* determination of the shape of the relationships between dependent and independent variables, and because the flexibility of the models accepts predictors that would violate assumptions of parametric approaches (Wood *et al.*, 2016).

Structural Equation Modeling (SEM) was used to explore how water depth influences potential biodiversity drivers (i.e., species richness, species dominance, species asynchrony, dominant species stability, common species stability, and rare species stability) of the temporal stability of community biomass. An initial model was constructed to include all potential pathways. (Figure Supplementary 3). The Fisher's C statistic and Akaike information criterion (AIC) were used to assess the fit of the model. The SEM analysis was done by using the restricted maximum-likelihood estimation and incorporating different lake bays as a random effect.

Statistical analyses were conducted in R (version 4.0.0., R development Core Team). The 'segmented' package was used to perform break-point analysis. The 'car' package was used to perform the analysis of covariance (ANCOVA). GAM models were fitted using the 'gam' function in the R package 'mgcv' (Wood 2011; Wood *et al.*, 2016). Partial regressions were conducted by using the R package 'effect'. The SEM analysis was performed with 'piecewiseSEM' package using the restricted maximum-likelihood estimation method and incorporating the random effect of block (Lefcheck 2016).

**Results**

*Community composition along the water depth gradient*

The biomass ratio of both common and rare species was much lower than that of dominant species across water depth gradients. The dominant, common, and rare species consisted of 3, 6, and 10 species, accounting for 79.1%, 19.4%, and 1.6% of community biomass, respectively (Table Supplementary 1). Along the water depth gradient, the biomass of *P. maackianus* showed a single-peak pattern, the relative biomass ratio of *C. demersum* showed a U-shaped pattern, and the relative biomass ratio of *V. natans* increased with the water depth. The biomass of most common and rare species decreased with water depth (Figure 2).

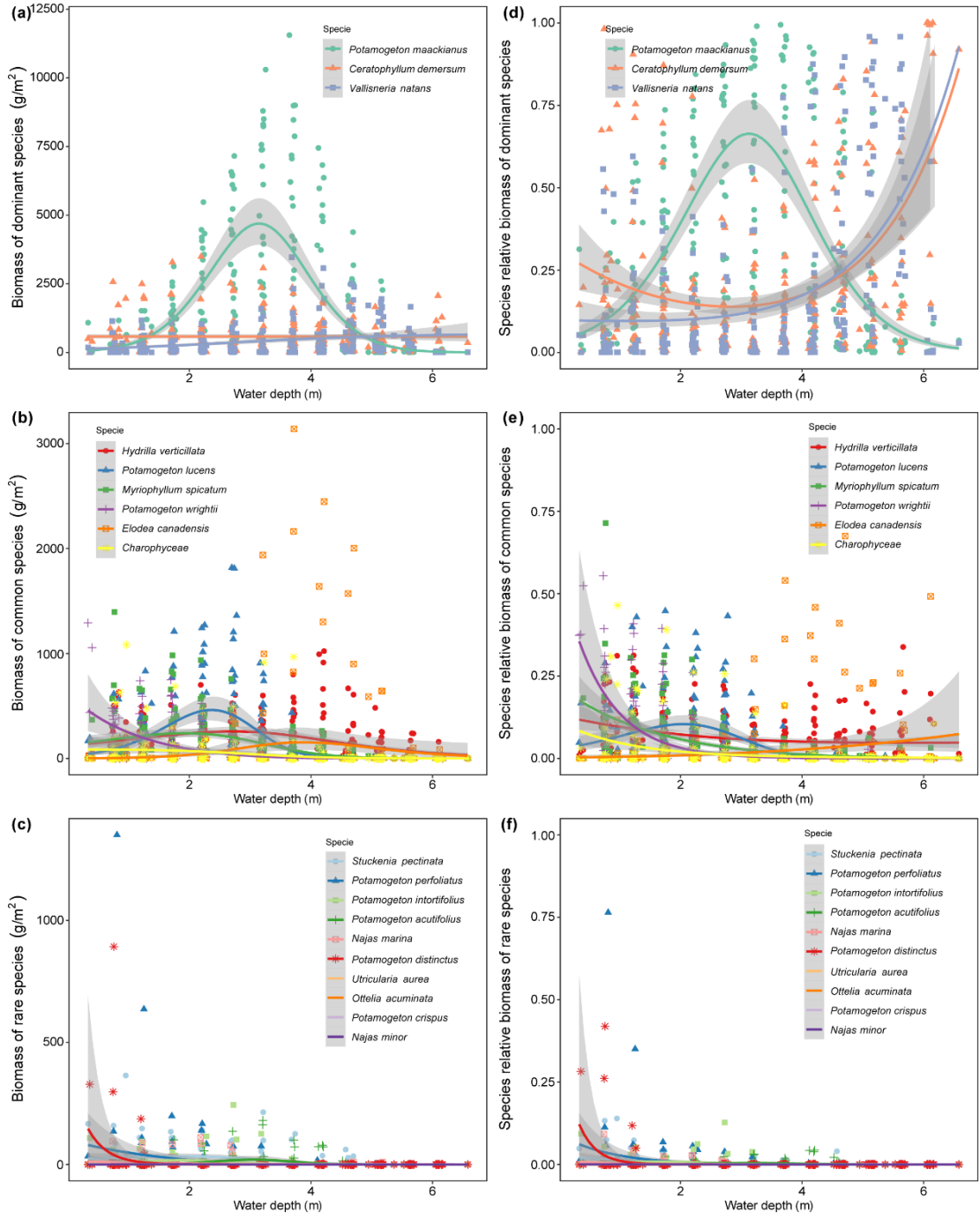

**Figure 2**. Biomass of different species along the water depth gradient. (a) The biomass of dominant species. (b) The biomass of common species. (c) The biomass of rare species. (d) Species relative abundance of dominant species. (e) Species relative abundance of common species. (f) Species relative abundance of rare species. Different colors represent different species. The scale is 0–1, i.e., 0–100%. Each dot represents community biomass at each water depth in each transect of each bay. The curves in all panels (a-f) were obtained by fitting generalized additive models.

*Temporal stability of the Community along the water depth gradient*

Using a segmented regression model to explore the community biomass stability along the water depth gradient, the biomass stability of the community showed a unimodal relationship along the water depth gradient ($p < 0.001$, Figure 3). In order to distinguish the two water depth areas (i.e. shallow water and deep water) of community biomass stability, we applied a break-point analysis within a piecewise linear regression analysis. The result showed that the largest value of community biomass stability was at a depth of 3.1 m ($p < 0.001$, Figure 3).

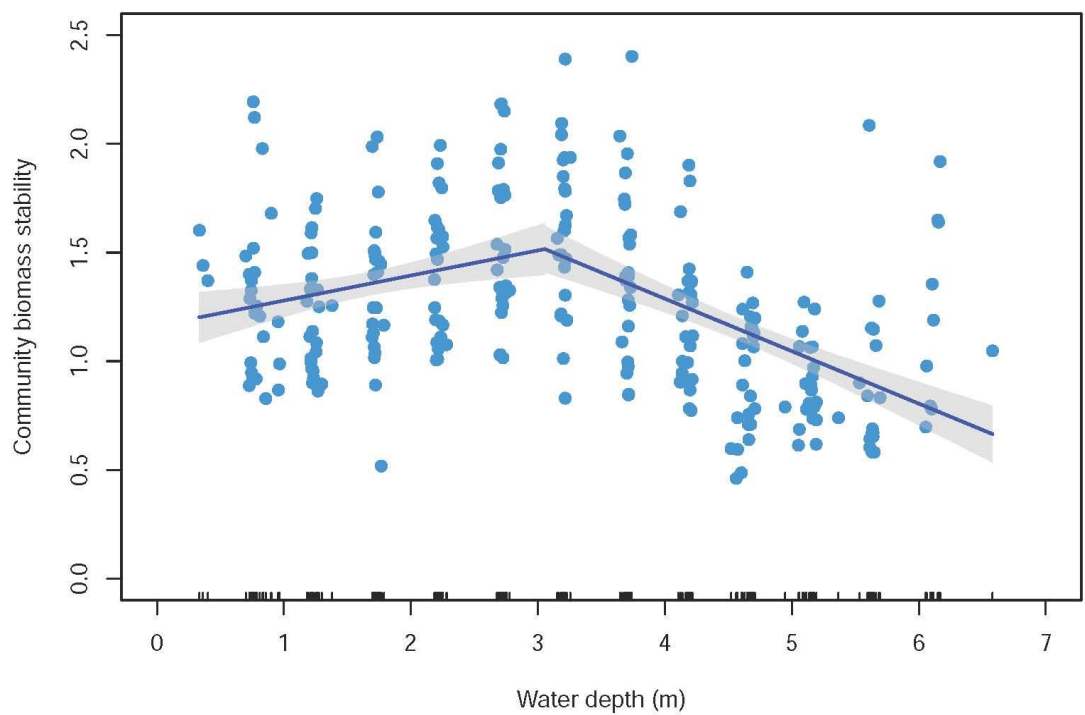

**Figure 3**. Relationship between the biomass stability of the community and water depth. The blue solid line represents the two-segment piecewise linear regression between water depth and community biomass stability based on the two-segment piecewise linear regression fit, and the shaded regions indicate the 95% confidence intervals of predictions.

*Drivers of stability along the water depth gradient*

Species richness and species asynchrony and dominant species stability both showed unimodal relationships with increasing water depth (*p* < 0.001; Figure 4a-d); The stability of common species and rare species decreased significantly with water depth (*p* < 0.001; Figure 4e-f); There was a significant positive correlation between species dominance and water depth (*p* < 0.001; Figure 4b).

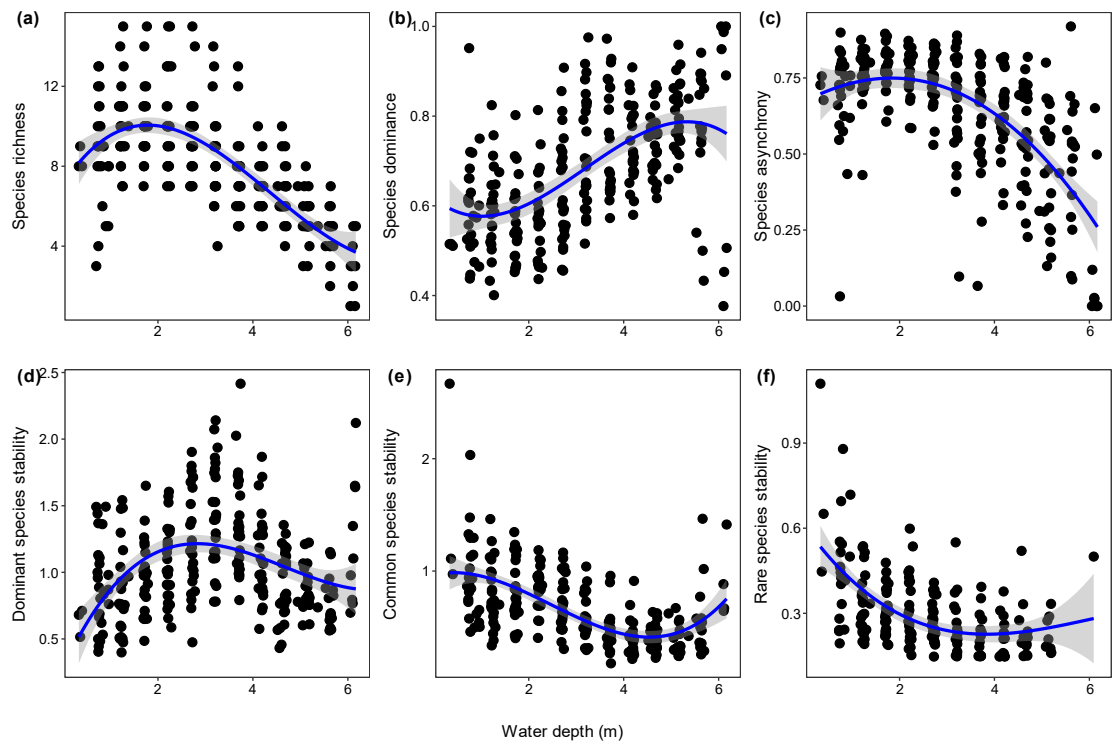

**Figure 4**. Relationships between water depth and main drivers of community stability. (a) Species richness; (b) species asynchrony; (c) species dominance (based on Simpson index); (d) dominant species stability; (e) common species; (f) rare species. The blue solid lines represent the fitted generalized additive models, and the shaded areas represent 95% confidence intervals of predictions. Each dot represents the temporal stability of the community biomass at each water depth in each bay. The curves in panels (a-f) were obtained by fitting generalized additive models.

*Effects of water depth on mean-variance scaling*

The water depth was first divided into shallow and deep areas based on break-point analyses. Following this, we observed that log-transformed values of the species variance in biomass were positively correlated to the log-transformed values of its mean values within both shallow and deep water areas, respectively (Figure 5; both $p < 0.001$). The slopes of the scaling relationship Z for both the shallow and deep-water areas were $1.62 \pm 0.01$ and $1.65 \pm 0.01$ (Mean ± SE), respectively, indicating that species richness is expected to promote the biomass stability of the plant community via the mean-variance scaling in both lake zones. Moreover, ANCOVA analyses showed that there is a significant change in the scaling of the coefficient Z between different water depth areas ($F = 4.6$, df = 1, $p < 0.05$).

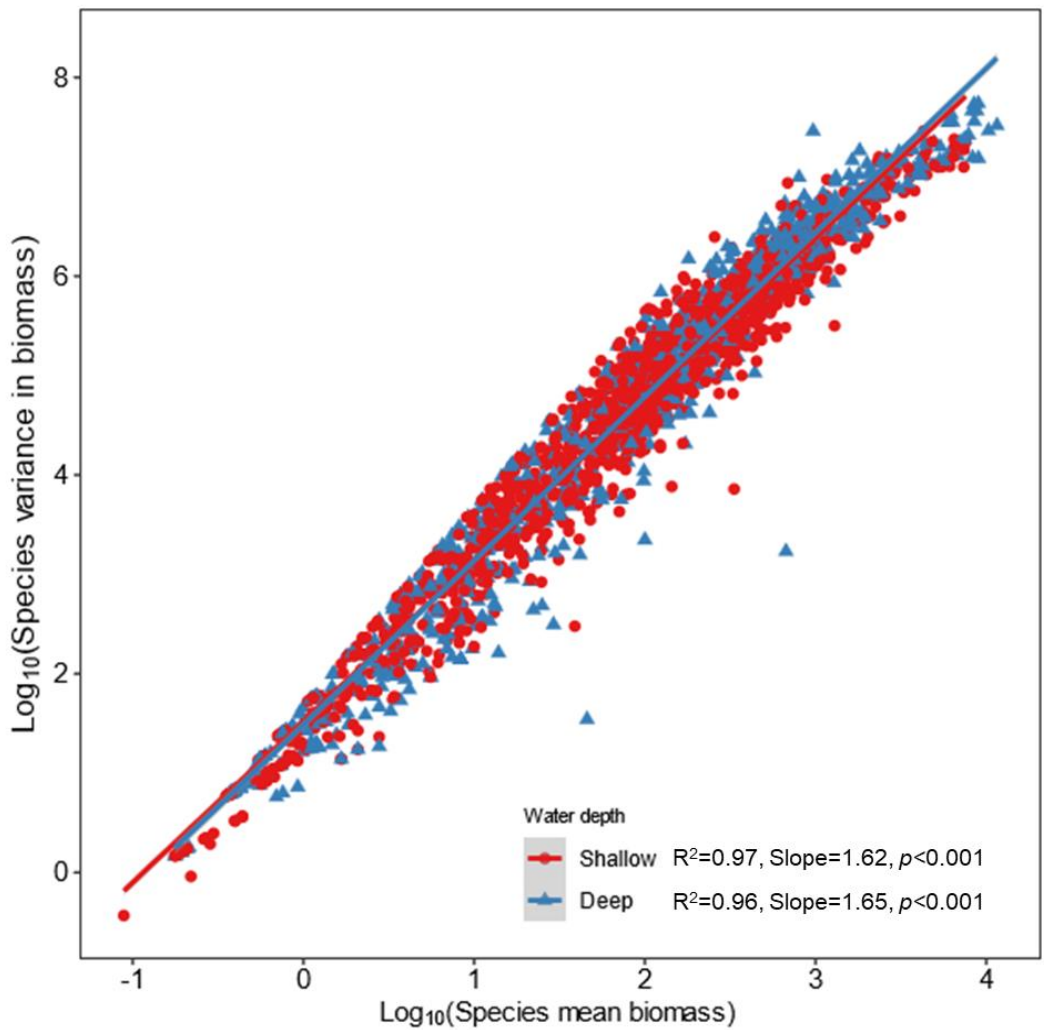

**Figure 5**. Relationship between the logarithm of variance in community biomass and the logarithm of the mean biomass per 50 m$^2$ sampling area at each water depth and for each species. The water depth was divided into shallow and deep areas based on break-

point analyses (see Figure 2). "Z" represents the slope of scaling coefficients.

*Pathways influencing temporal stability along the water depth gradient*

Structural Equation Modelling (SEM) analysis showed multiple pathways by which the temporal stability of the community increased along the water depth gradient. The stability of dominant species and species asynchrony played a positive role in determining the temporal stability of the community biomass within both shallow and deep areas. However, the relationship between water depth and drivers of the temporal stability of the community changed from positive to negative in shallow and deep areas, respectively, indicating the effects of water depth on the temporal stability of the community biomass changed from stabilizing to destabilizing with increasing water depth within each lake zone (Figure 6a and 6b).

Within shallow waters, the effect of water depth on the biomass stability of the plant community was positive, with the standardized total effect size being 0.25 (Figure 6a and Table Supplementary 2). While water depth significantly increased the biomass stability of dominant species (path efficient = 0.36, $p < 0.001$), it reduced the biomass stability of common species (path efficient = -0.25, $p < 0.001$) and rare species (path efficient = -0.43, $p < 0.001$). Species dominance, species asynchrony, and biomass stability of dominant species are all important pathways that changed with water depth in the way they contributed to the temporal stability of biomass at the community level. In contrast, the temporal stability of the biomass of common species decreased the biomass stability at the community level, an effect that was reduced with increasing water depth. Moreover, species richness was positively associated with plant community biomass stability in both shallow and deep waters (Figure 6a and 6b). The SEM (with bay as a random effect in shallow waters) explained 85% of the variation in the temporal stability of plant community biomass (Figure 6a).

In contrast, within deep waters, the effect of water depth on plant community biomass stability was negative, with the standardized total effect size being -0.55 (Figure 6b and Table Supplementary 3). Water depth has a negative effect on species asynchrony and the temporal stability of the biomass of the dominant species (Figure

6b). Species asynchrony and dominant species still showed a positive association with the temporal stability of community biomass (Figure 6b). In addition, species richness declined sharply with increasing water depth. The SEM (with bay as a random effect in deep waters) explained 96% of the variation in the temporal stability of plant community biomass (Figure 6b).

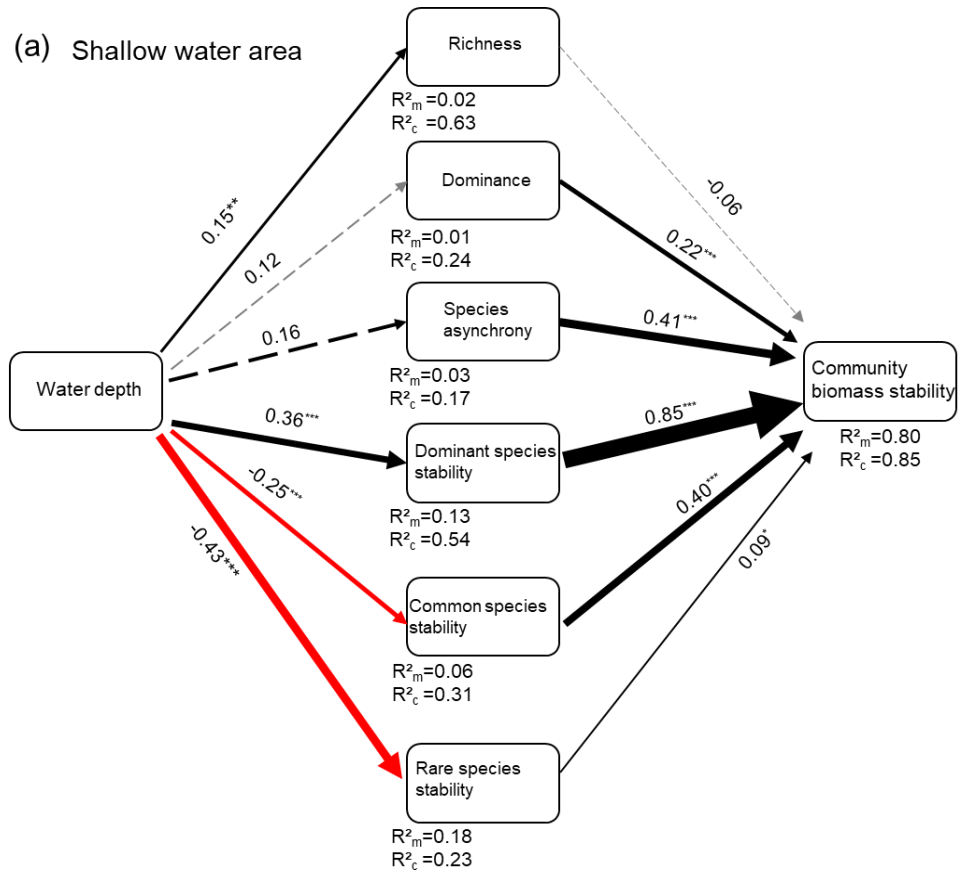

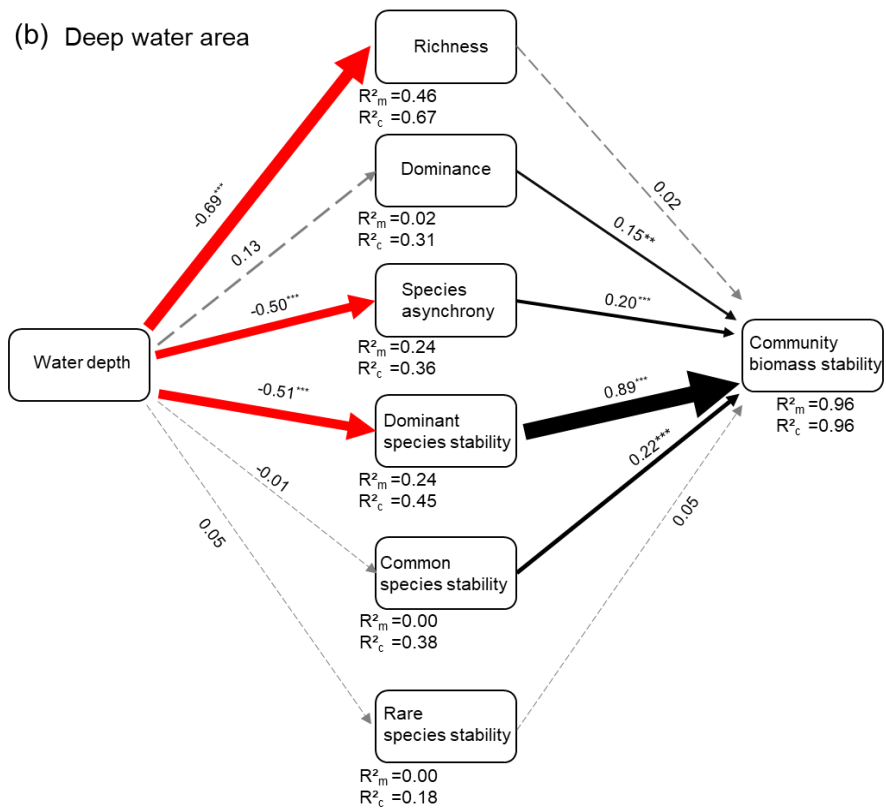

**Figure 6**. Effect of multiple drivers of temporal stability as a function of water depth, evaluated using the Structural Equation Modelling (SEM) analysis. Black and red arrows indicate significant positive and negative pathways, respectively, and grey dashed arrows indicate non-significant pathways. Bold numbers represent the standard path coefficients. The arrow width is proportional to the strength of the relationship. $R_m^2$ and $R_c^2$ are the variance explained by all paths from the fixed effects and combinations from the fixed and random effects, respectively. For the shallow water depth area: Fisher C = 0.396, $p$ = 0.82, AIC = -28.289. For the deep water depth area: Fisher C = 2.466, $p$ = 0.291, AIC = 36.426. Level of significance: ***$p$ < 0.001; **0.001 ≤ $p$ < 0.01; *0.01 ≤ $p$ < 0.05.

**Discussion**

This study demonstrated that environmental factors can have oppositive (i.e., but not limited to either positive or negative) influences on the temporal stability of the community by changing the drivers of biodiversity (e.g., dominant species biomass stability and species asynchrony). For example, our field study found that although biodiversity drivers always positively contributed to the temporal stability of the submerged plant communities in both shallow and deep water, water depth shifted the temporal stability from a positive stabilizing effect in shallow waters to a negative destabilizing effect in deep waters.

*Unimodal response of community stability along the water depth gradient*

Our field survey found that the stability of the community biomass of submerged macrophytes showed a unimodal pattern with the full water depth, and with the highest values at an intermediate water depth. We found that biodiversity drivers (i.e., dominant species stability, species asynchrony, and species richness) jointly contribute to the temporal stability of the submerged plant communities in both shallow and deep water. It is worth noting that all these biodiversity drivers (i.e., dominant species stability, species asynchrony, and species richness) showed a unimodal pattern along the water depth gradient.

Previous studies have also found that species richness, species asynchrony, and the stability of dominant species show nonlinear patterns along environmental gradients, and both positive and negative relationship patterns were found (Garcia-Palacios *et al.*, 2018; Zhang *et al.*, 2018). For example, there are both positive and negative relationship patterns exist along aridity gradients (Garcia-Palacios *et al.*, 2018). Also, increased climate variability has been shown to increase species diversity and further increase species asynchrony (P Chesson *et al.*, 1989). However, high levels of climate variability can also reduce species diversity (Peter B Adler *et al.*, 2008; Shurin *et al.*, 2010) and can lead to a decrease in species asynchrony, and ultimately a decrease in community stability (Loreau & de Mazancourt 2013; Gilbert *et al.*, 2020). Thus, the effect of water depth on the temporal stability of the community essentially depends on whether these biodiversity drivers are at the beginning or end of the water depth gradient.

*Effect of water depth on stabilizing mechanisms of community stability in different water depth ranges*

Despite the opposite environmental effects that occur in the shallow water and deep water plots, the stabilization of dominant species generally contributes to the stabilization of the macrophyte communities. Our results demonstrated that the stability of dominant species was an important driver for the stability of community biomass (Figure 6), which was consistent with the results of previous studies (Xu *et al.*, 2015; Ma *et al.*, 2017; Ma *et al.*, 2021; Quan *et al.*, 2021).

In both shallow and deep water during the 4-year field study, water depth enhanced or weakened the biomass stability of the plant community by increasing or decreasing the stability of the dominant species stability, respectively. Specifically, three dominant macrophyte species (*P. maackianus, V. natans, and C. demersum*) accounted for 79.1% of the community biomass (Table Supplementary 1). In shallow water area in this study, the space limitation is the main limiting factor, First, the plant height is positively correlated with the water depth (Fu *et al.*, 2014; Fu *et al.*, 2018; He *et al.*, 2019). Also, the branch length increases with the increase of the volume of the water column volume, which improves its ability to occupy more space and reduce competition for space(Fu *et al.*, 2014; Fu *et al.*, 2018; He *et al.*, 2019). As a result, the stability of the dominant species (e.g., *P. maackianus*) of submerged plants increases with increases in the water depth (Fu *et al.*, 2014; Fu *et al.*, 2018; He *et al.*, 2019). In contrast, in deep water, the low availability of light becomes the main limiting factor, and the total leaf area and leaf area ratio of submerged macrophytes decrease. This results in the decrease of photosynthetic biomass accumulation. Consequently, the stability of the dominant species decreases with the increase of water depth (Fu *et al.*, 2014; Fu *et al.*, 2018; He *et al.*, 2019).

Our study also demonstrated that species asynchrony is an important mechanism that modulates community stability, which is consistent with the results of previous studies in terrestrial ecosystems (Xu *et al.*, 2015; Zhang *et al.*, 2016; Ma *et al.*, 2017; Quan *et al.*, 2021). For example, in both shallow and deep water, species asynchrony was consistently, positively correlated with the stability of the community biomass (Figure 5). Within the shallow water depth area, there was sufficient light availability that increases interspecific competition and results in higher species asynchrony. In addition, higher species richness often promotes higher differentiation of temporal niches, thus leading to higher species asynchrony within the shallow water area. In contrast, within deep waters, water depth reduces community stability by weakening species asynchrony.

Because of reduced underwater light availability in deep water area (Figure Supplementary 4), most species are subjected to low light availability (Figure 2). Species turnover occurs mainly in shallow water, because common and rare species, which occupy a higher proportion of species richness, are mainly concentrated in shallow waters, the number of species in deep water is reduced, as no new species can be established. For example, previous studies found that the response strategy of submerged macrophytes to low light conditions largely depends on their growth form. The canopy-forming species (e.g., *P. maackianus*) respond to low light conditions at deeper sites by stem elongation. In contrast, the rosette-type species (e.g., *V. natans*) increase the chlorophyll content of their shoots (He *et al.*, 2019). Hence, only the dominant species *C. demersum* and *V. natans* had a relatively high abundance in areas above 4.5m depth, and these are followed by the common species of *H. verticillata* and *E. nuttallii* (Figure 2). These low light tolerant species have similar life-cycle rhythms in deep waters, thus resulting in reduced asynchrony (Wen *et al.*, 2022).

In both shallow and deep water, where the results of the environmental effects were opposite, the stability of common species contributed to increased stability at the community level but this effect was weakened in deep waters. For example, common species represented 31.6% of species richness and accounted for 19.4% of plant community biomass. Rare species represented 52.6% of species richness, however they accounted for only 0.3% of plant community biomass (Table Supplementary 1).

Common and rare species had significant effects on community biomass stability, which has rarely been reported in previous studies (Douda *et al.*, 2018; Ma *et al.*, 2021; Zong *et al.*, 2023). However, these results are in accordance to findings found in Ma *et al.* (2021) who found that at low N addition rates, the stability of common species had a significant, positive correlation with the temporal stability of the community. However, this relationship became insignificant at high rates of nitrogen additions. In our study, with increasing water depth in the shallow-water depth area, common species and rare species significantly decreased because of the relatively higher levels of competition at intermediate water depths. This effect was amplified, and even led to species extinction in the deep-water depth area because of stronger low-light limitation. Therefore, the contribution of the stability of common and rare species to temporal stability at the community level cannot be ignored under certain environmental ranges.

Our results that were based on structural equation models suggest that species richness is not a significant predictor of the stability of the biomass of plant communities. Numerous theoretical and empirical studies have shown that increasing species diversity can enhance community stability (Tilman *et al.*, 2006; Jiang & Pu 2009; Hector *et al.*, 2010; Hautier *et al.*, 2014; Quan *et al.*, 2021). However, previous studies have also shown that other biodiversity drivers can have stronger influences (e.g. species asynchrony and temporal stability of dominant species) on the temporal stability of the community than that of species richness (Xu *et al.*, 2015; Valencia *et al.*, 2020; Ma *et al.*, 2021). In deep water in our study, although species richness decreased significantly with increasing depth, we only found changes in common and rare species. Alternatively, because the mean-variance relationship is also considered as an universal mechanism driving community biomass stability, communities with more species are more likely to reduce the variance of ecosystem attributes through statistical averaging (Doak *et al.*, 1998). Moreover, factors altering the slope of such a relationship can change ecosystem stability. In this study, the Z value of the slope is between 1 and 2, which is expected to increase community stability within both water depth areas (Tilman *et al.*, 1998). However, the increase in dominance of certain species (and therefore a reduction in evenness) with increasing water depth may lead to a weaker effect of species richness on the stability of community biomass (Cottingham *et al.*, 2001). Hence, the mean-variance scaling is likely negligible in this study because the portfolio effect is generally stronger in communities with species biomass that is more evenly distributed (Doak *et al.*, 1998; Hillebrand *et al.*, 2008; Grman *et al.*, 2010).

*Management Implications and Conclusions*

Our study results suggest that the direction of environmental change on the temporal stability of the community depends on its effect on biodiversity drivers. Our study may provide insights into how community stability varies along large environmental gradients at the local scale. For example, we found that along relatively large environmental gradients (such as the shallow water and deep waters covered in this study), the local environmental condition (e.g., water depth) can influence the temporal stability of the community in aquatic ecosystems in opposite ways when biodiversity promotes temporal stability of the community, the local environment can act as both stabilizing and destabilizing drivers along environment gradients by opposingly influence the dominant species stability and species asynchrony. Consequently, our findings have implications for the protection and management of ecosystems that attempt to conserve the stability of the functions and services of the natural ecosystem.

Our study provides a perspective to better understand how the relationships among biodiversity, stability, and environmental change have important implications for ecosystem management and protection. Our results suggest that the stabilization mechanisms operating along different environmental ranges should be considered in planning ecosystem conservation and restoration. Specifically, changes in environmental factors may promote or decrease community stability under different environmental ranges. For example, in lake restoration, increasing species diversity and dominant species in shallow water can improve the stability of submerged plant communities. However, in deep water, the addition of species that can withstand low light conditions can reduce community species diversity and species asynchrony, both of which are conducive to maintaining community stability.

# References

Chaudhary, C., Richardson, A. J., Schoeman, D. S. & Costello, M. J. (2021). Global warming is causing a more pronounced dip in marine species richness around the equator. Proceedings of the National Academy of Sciences of the United States of America, 118.

Cottingham, K. L., Brown, B. L. & Lennon, J. T. (2001). Biodiversity may regulate the temporal variability of ecological systems. Ecology Letters, 4, 72-85.

Doak, D. F., Bigger, D., Harding, E. K., Marvier, M. A., O'Malley, R. E. & Thomson, D. (1998). The statistical inevitability of stability-diversity relationships in community ecology. The American Naturalist, 151, 264-276.

Dodds, W. K., Bruckerhoff, L., Batzer, D., Schechner, A., Pennock, C., Renner, E., Tromboni, F., Bigham, K. & Grieger, S. (2019). The freshwater biome gradient framework: predicting macroscale properties based on latitude, altitude, and precipitation. Ecosphere, 10.

Douda, J., Doudova, J., Hulik, J., Havrdova, A. & Boublik, K. (2018). Reduced competition enhances community temporal stability under conditions of increasing environmental stress. Ecology, 99, 2207-2216.

Fu, H., Yuan, G., Lou, Q., Dai, T., Xu, J., Cao, T., Ni, L., Zhong, J. & Fang, S. (2018). Functional traits mediated cascading effects of water depth and light availability on temporal stability of a macrophyte species. Ecological Indicators, 89, 168-174.

Fu, H., Yuan, G., Ozkan, K., Johansson, L. S., Sondergaard, M., Lauridsen, T. L. & Jeppesen, E. (2021). Patterns of Seasonal Stability of Lake Phytoplankton Mediated by Resource and Grazer Control During Two Decades of Re-oligotrophication. Ecosystems, 24, 911-925.

Fu, H., Zhong, J., Yuan, G., Xie, P., Guo, L., Zhang, X., Xu, J., Li, Z., Li, W., Zhang, M., Cao, T. & Ni, L. (2014). Trait-based community assembly of aquatic macrophytes along a water depth gradient in a freshwater lake. Freshwater Biology, 59, 2462-2471.

Garcia-Palacios, P., Gross, N., Gaitan, J. & Maestre, F. T. (2018). Climate mediates the biodiversity-ecosystem stability relationship globally. Proceedings of the National Academy of Sciences of the United States of America, 115, 8400-8405.

Gilbert, B., MacDougall, A. S., Kadoya, T., Akasaka, M., Bennett, J. R., Lind, E. M., Flores-Moreno, H., Firn, J., Hautier, Y., Borer, E. T., Seabloom, E. W., Adler, P. B., Cleland, E. E., Grace, J. B., Harpole, W. S., Esch, E. H., Moore, J. L., Knops, J., McCulley, R., Mortensen, B., Bakker, J., Fay, P. A. & Kerkhoff, A. (2020). Climate and local environment structure asynchrony and the stability of primary production in grasslands. Global Ecology and Biogeography, 29, 1177-1188.

Gonzalez, A. & Loreau, M. (2009). The Causes and Consequences of Compensatory Dynamics in Ecological Communities. Annual Review of Ecology Evolution and Systematics, 40, 393-414.

Grime, J. P. (1998). Benefits of plant diversity to ecosystems: immediate, filter and founder effects. Journal of Ecology, 86, 902-910.

Grman, E., Lau, J. A., Schoolmaster Jr., D. R. & Gross, K. L. (2010). Mechanisms contributing to stability in ecosystem function depend on the environmental context. Ecology Letters, 13, 1400-1410.

Hautier, Y., Seabloom, E. W., Borer, E. T., Adler, P. B., Harpole, W. S., Hillebrand, H., Lind, E. M., MacDougall, A. S., Stevens, C. J., Bakker, J. D., Buckley, Y. M., Chu, C., Collins, S. L., Daleo,

P., Damschen, E. I., Davies, K. F., Fay, P. A., Firn, J., Gruner, D. S., Jin, V. L., Klein, J. A., Knops, J. M. H., La Pierre, K. J., Li, W., McCulley, R. L., Melbourne, B. A., Moore, J. L., O'Halloran, L. R., Prober, S. M., Risch, A. C., Sankaran, M., Schuetz, M. & Hector, A. (2014). Eutrophication weakens stabilizing effects of diversity in natural grasslands. Nature, 508, 521-525.

He, L., Zhu, T., Wu, Y., Li, W., Zhang, H., Zhang, X., Cao, T., Ni, L. & Hilt, S. (2019). Littoral Slope, Water Depth and Alternative Response Strategies to Light Attenuation Shape the Distribution of Submerged Macrophytes in a Mesotrophic Lake. Frontiers in Plant Science, 10, 169.

Hector, A., Hautier, Y., Saner, P., Wacker, L., Bagchi, R., Joshi, J., Scherer-Lorenzen, M., Spehn, E. M., Bazeley-White, E., Weilenmann, M., Caldeira, M. C., Dimitrakopoulos, P. G., Finn, J. A., Huss-Danell, K., Jumpponen, A., Mulder, C. P. H., Palmborg, C., Pereira, J. S., Siamantziouras, A. S. D., Terry, A. C., Troumbis, A. Y., Schmid, B. & Loreau, M. (2010). General stabilizing effects of plant diversity on grassland productivity through population asynchrony and overyielding. Ecology, 91, 2213-2220.

Hillebrand, H., Bennett, D. M. & Cadotte, M. W. (2008). Consequences of dominance: A review of evenness effects on local and regional ecosystem processes. Ecology, 89, 1510-1520.

Hooper, D. U., Chapin, F. S., Ewel, J. J., Hector, A., Inchausti, P., Lavorel, S., Lawton, J. H., Lodge, D. M., Loreau, M., Naeem, S., Schmid, B., Setala, H., Symstad, A. J., Vandermeer, J. & Wardle, D. A. (2005). Effects of biodiversity on ecosystem functioning: A consensus of current knowledge. Ecological Monographs, 75, 3-35.

Hooper, D. U., Chapin, F. S., Ewel, J. J., Hector, A., Inchausti, P., Lavorel, S., Lawton, J. H., Lodge, D. M., Loreau, M., Naeem, S., Schmid, B., Setälä, H., Symstad, A. J., Vandermeer, J. & Wardle, D. A. (2005). Effects of biodiversity on ecosystem functioning: A consensus of current knowledge. Ecological Monographs, 75, 3-35.

Jiang, L. & Pu, Z. (2009). Different Effects of Species Diversity on Temporal Stability in Single-Trophic and Multitrophic Communities. The American Naturalist, 174, 651-659.

Lefcheck, J. S. (2016). piecewiseSEM: Piecewise structural equation modelling in r for ecology, evolution, and systematics. Methods in Ecology and Evolution, 7, 573-579.

Leps, J. (2004). Variability in population and community biomass in a grassland community affected by environmental productivity and diversity. Oikos, 107, 64-71.

Lewerentz, A., Hoffmann, M. & Sarmento Cabral, J. (2021). Depth diversity gradients of macrophytes: Shape, drivers, and recent shifts. Ecology and Evolution, 11, 13830-13845.

Loreau, M. & de Mazancourt, C. (2013). Biodiversity and ecosystem stability: a synthesis of underlying mechanisms. Ecology Letters, 16 Suppl 1, 106-15.

Ma, F., Zhang, F., Quan, Q., Song, B., Wang, J., Zhou, Q. & Niu, S. (2021). Common Species Stability and Species Asynchrony Rather than Richness Determine Ecosystem Stability Under Nitrogen Enrichment. Ecosystems, 24, 686-698.

Ma, Z., Liu, H., Mi, Z., Zhang, Z., Wang, Y., Xu, W., Jiang, L. & He, J. S. (2017). Climate warming reduces the temporal stability of plant community biomass production. Nature Communications, 8, 15378.

Middelboe, A. L. & Markager, S. (1997). Depth limits and minimum light requirements of freshwater macrophytes. Freshwater Biology, 37, 553-568.

Muggeo, V. M. (2003). Estimating regression models with unknown break-points. Statistics in

Medicine, 22, 3055-71.

Ouyang, S., Xiang, W., Gou, M., Chen, L., Lei, P., Xiao, W., Deng, X., Zeng, L., Li, J., Zhang, T., Peng, C. & Forrester, D. I. (2021). Stability in subtropical forests: The role of tree species diversity, stand structure, environmental and socio-economic conditions. Global Ecology and Biogeography, 30, 500-513.

P Chesson, nbsp & Huntly, N. (1989). Short-term instabilities and long-term community dynamics. Trends in Ecology & Evolution, 4 10, 293-8.

Peter B Adler, nbsp & Drake, J. M. (2008). Environmental variation, stochastic extinction, and competitive coexistence. The American Naturalist, 172 5, 186-95.

Peters, M. K., Hemp, A., Appelhans, T., Behler, C., Classen, A., Detsch, F., Ensslin, A., Ferger, S. W., Frederiksen, S. B., Gebert, F., Haas, M., Helbig-Bonitz, M., Hemp, C., Kindeketa, W. J., Mwangomo, E., Ngereza, C., Otte, I., Röder, J., Rutten, G., Schellenberger Costa, D., Tardanico, J., Zancolli, G., Deckert, J., Eardley, C. D., Peters, R. S., Rödel, M.-O., Schleuning, M., Ssymank, A., Kakengi, V., Zhang, J., Böhning-Gaese, K., Brandl, R., Kalko, E. K. V., Kleyer, M., Nauss, T., Tschapka, M., Fischer, M. & Steffan-Dewenter, I. (2016). Predictors of elevational biodiversity gradients change from single taxa to the multi-taxa community level. Nature Communications, 7, 13736.

Quan, Q., Zhang, F., Jiang, L., Chen, H. Y. H., Wang, J., Ma, F., Song, B. & Niu, S. (2021). High-level rather than low-level warming destabilizes plant community biomass production. Journal of Ecology, 109, 1607-1617.

Quintero, I. & Jetz, W. (2018). Global elevational diversity and diversification of birds. Nature, 555, 246-250.

R Core Team. 2021. R: A language and environment for statistical computing. Vienna, Austria: R Foundation for Statistical Computing. https://www.R-project.org/.

Shurin, J. B., Winder, M., Adrian, R., Keller, W., Matthews, B., Paterson, A. M., Paterson, M. J., Pinel-Alloul, B., Rusak, J. A. & Yan, N. D. (2010). Environmental stability and lake zooplankton diversity – contrasting effects of chemical and thermal variability. Ecology Letters, 13, 453-463.

Spence, D. H. N. (1982). The Zonation of Plants in Freshwater Lakes. Advances in Ecological Research.

Tilman, D., Lehman, C. L. & Bristow, C. E. (1998). Diversity-stability relationships: statistical inevitability or ecological consequence? The American Naturalist, 151, 277-82.

Tilman, D., Reich, P. B. & Knops, J. M. (2006). Biodiversity and ecosystem stability in a decade-long grassland experiment. Nature, 441, 629-32.

Valencia, E., de Bello, F., Galland, T., Adler, P. B., Leps, J., A, E. V., van Klink, R., Carmona, C. P., Danihelka, J., Dengler, J., Eldridge, D. J., Estiarte, M., Garcia-Gonzalez, R., Garnier, E., Gomez-Garcia, D., Harrison, S. P., Herben, T., Ibanez, R., Jentsch, A., Juergens, N., Kertesz, M., Klumpp, K., Louault, F., Marrs, R. H., Ogaya, R., Onodi, G., Pakeman, R. J., Pardo, I., Partel, M., Peco, B., Penuelas, J., Pywell, R. F., Rueda, M., Schmidt, W., Schmiedel, U., Schuetz, M., Skalova, H., Smilauer, P., Smilauerova, M., Smit, C., Song, M., Stock, M., Val, J., Vandvik, V., Ward, D., Wesche, K., Wiser, S. K., Woodcock, B. A., Young, T. P., Yu, F. H., Zobel, M. & Gotzenberger, L. (2020). Synchrony matters more than species richness in plant community stability at a global scale. Proceedings of the National Academy of Sciences of the United States of America, 117, 24345-24351.

Wen, Z., Wang, H., Zhang, Z., Cao, Y., Yao, Y., Gao, X., Tian, Y., Su, H., Ni, L., Xiao, W., Cai, Q., Zhang, X. & Cao, T. (2022). Depth distribution of three submerged macrophytes under water level fluctuations in a large plateau lake. Aquatic Botany, 176.

Wen, Z. H., Ma, Y. W., Wang, H., Cao, Y., Yuan, C. B., Ren, W. J., Ni, L. Y., Cai, Q. H., Xiao, W., Fu, H., Cao, T. & Zhang, X. L. (2021). Water level regulation for eco-social services under climate change in Erhai lake over the past 68 years in China. Frontiers in Environmental Science, 9.

Wood, S. N. (2011). Fast stable restricted maximum likelihood and marginal likelihood estimation of semiparametric generalized linear models. Journal of the Royal Statistical Society, 73, 3-36.

Wood, S. N., Pya, N. & Sfken, B. (2016). Smoothing parameter and model selection for general smooth models. JASA: Journal of the American Statistical Association, 111, 1548-1563.

Xu, Z., Ren, H., Li, M.-H., van Ruijven, J., Han, X., Wan, S., Li, H., Yu, Q., Jiang, Y. & Jiang, L. (2015). Environmental changes drive the temporal stability of semi-arid natural grasslands through altering species asynchrony. Journal of Ecology, 103, 1308-1316.

Yan, Y., Ma, F., Wang, J., Zhang, R., Peng, J., Liao, J., Zhou, Q. & Niu, S. (2023). Warming stabilizes alpine ecosystem facing extreme rainfall events by changing plant species composition. Journal of Ecology, 111, 2064-2076.

Ye, B., Chu, Z., Wu, A., Hou, Z. & Wang, S. (2018). Optimum water depth ranges of dominant submersed macrophytes in a natural freshwater lake. PLOS ONE, 13, e0193176.

Zhang, Y., Loreau, M., He, N., Wang, J., Pan, Q., Bai, Y. & Han, X. (2018). Climate variability decreases species richness and community stability in a temperate grassland. Oecologia, 188, 183-192.

Zhang, Y., Loreau, M., Lu, X., He, N., Zhang, G. & Han, X. (2016). Nitrogen enrichment weakens ecosystem stability through decreased species asynchrony and population stability in a temperate grassland. Global Change Biology, 22, 1445-55.

Zong, N., Hou, G., Shi, P. & Song, M. (2023). Winter warming alleviates the severely negative effects of nitrogen addition on ecosystem stability in a Tibetan alpine grassland. Science of the Total Environment, 855.

**Supplementary materials**

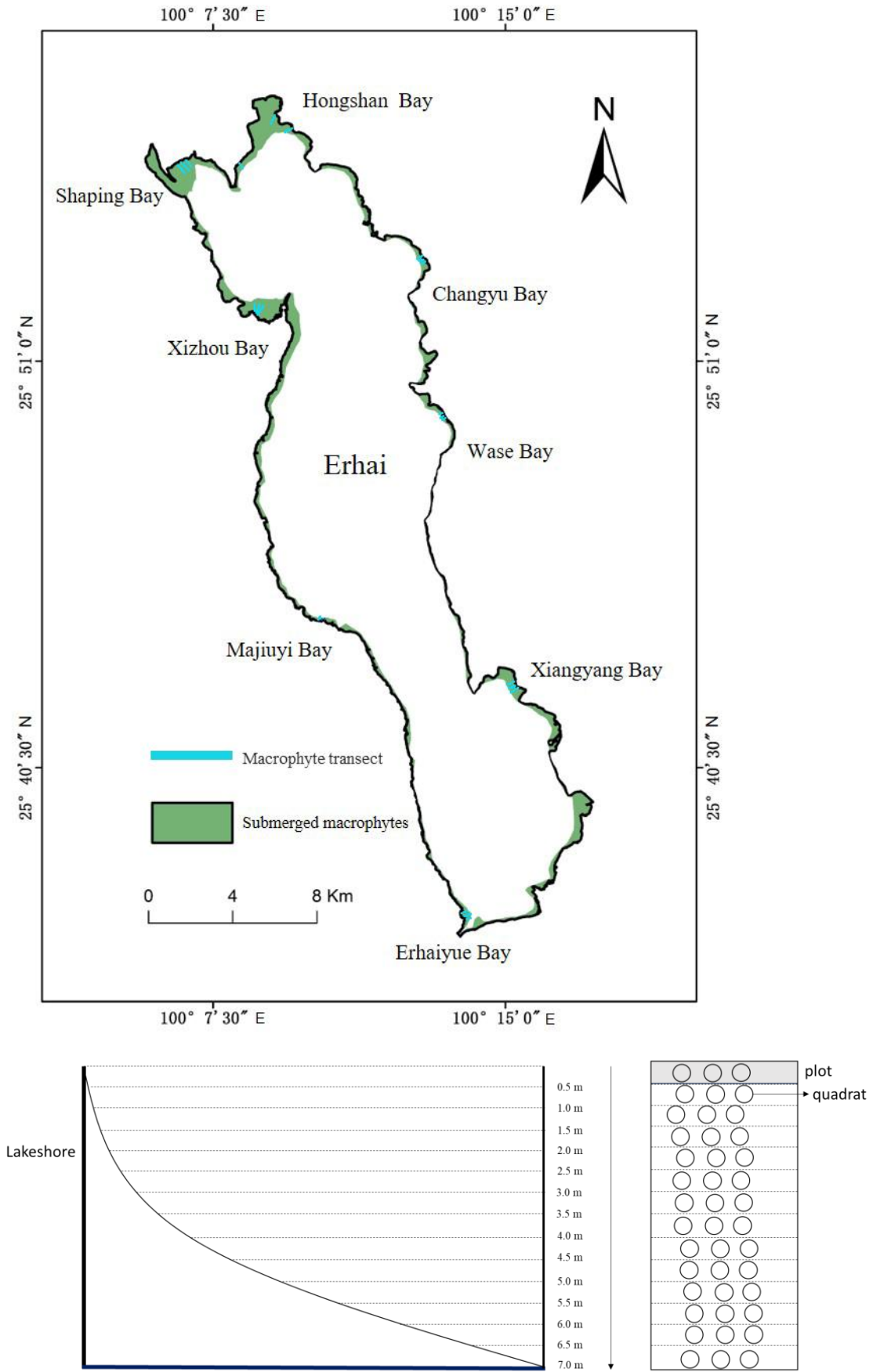

**Figure Supplementary 1.** Study location and sampling scheme. The study was carried

out in Erhan Lake, Yunnan, China. We sampled submerged macrophyte communities in 30,071 quadrats (0.2 m$^2$) in eight bays. Three permanent transects were established in each bay perpendicular to the lakeshore from shallow to deep water, with 5 × 5 m plots located along the water depth gradient at 0.5-m intervals from 0 to 7.0 m. Within each 25 m$^2$ plot, three 0.2 m$^2$ quadrats were used for analysis.

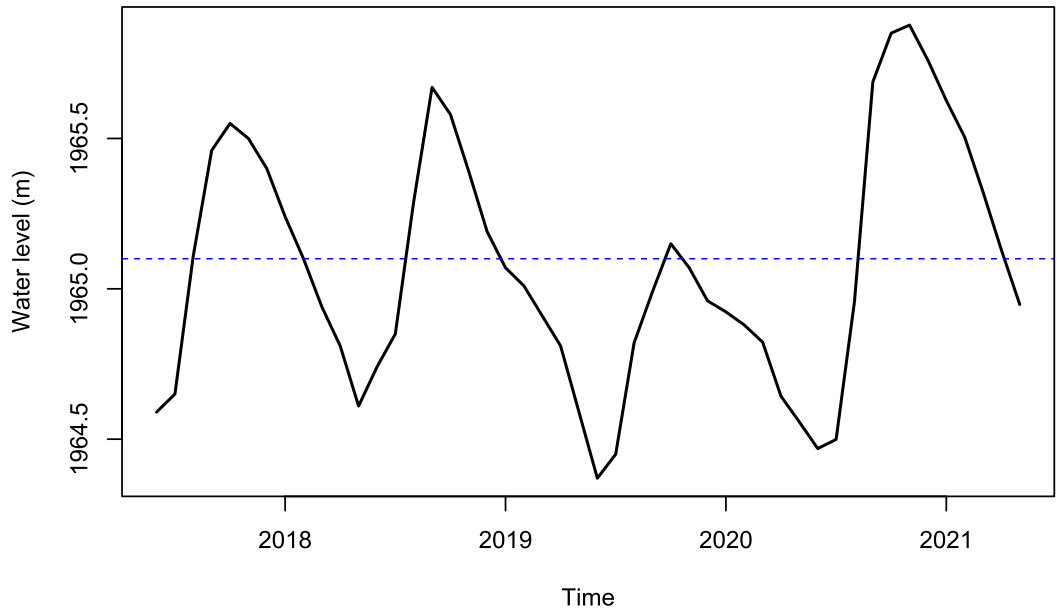

**Figure Supplementary 2**. Time series plot of water level fluctuations of Erha lake from June 2017 to May 2021. The blue dotted line represents the normal (mean) water level during this period was 1965.1m above sea level (a.m.).

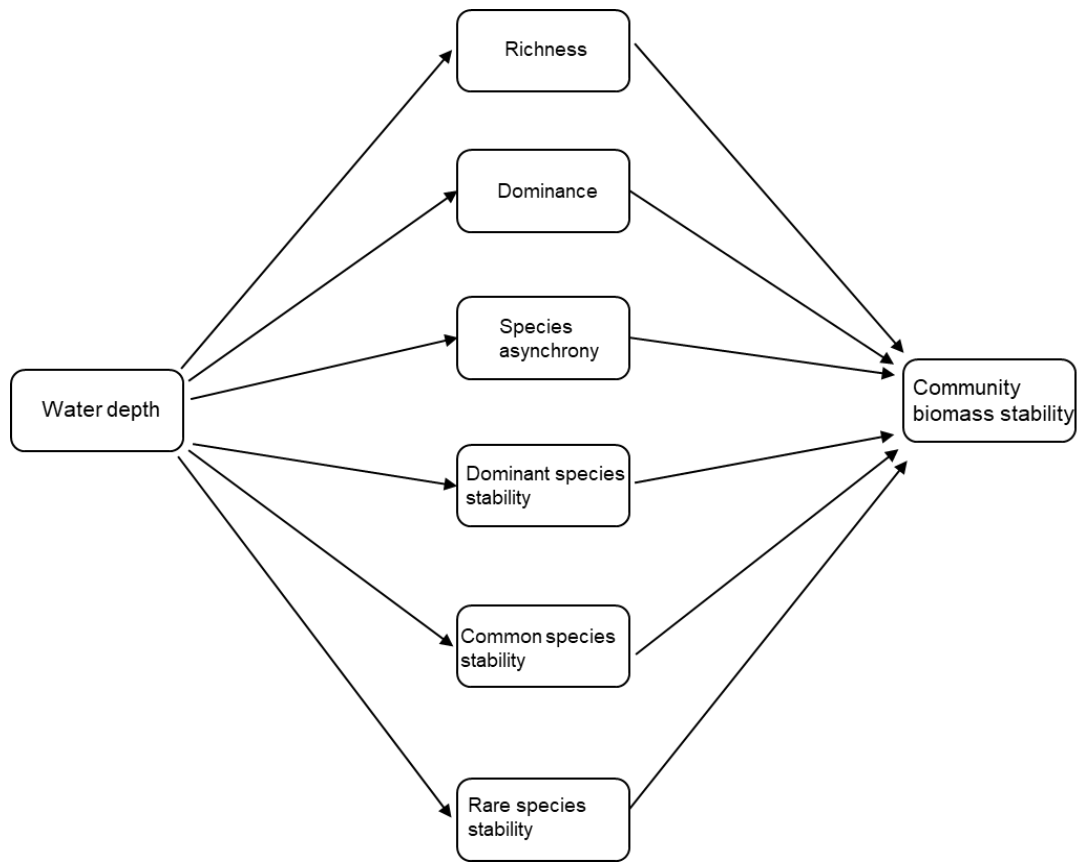

**Figure Supplementary 3**. The initial structural equation model for plant community biomass stability in response to water depth gradients. All plausible pathways were considered, on the basis of theoretical and empirical predictions.

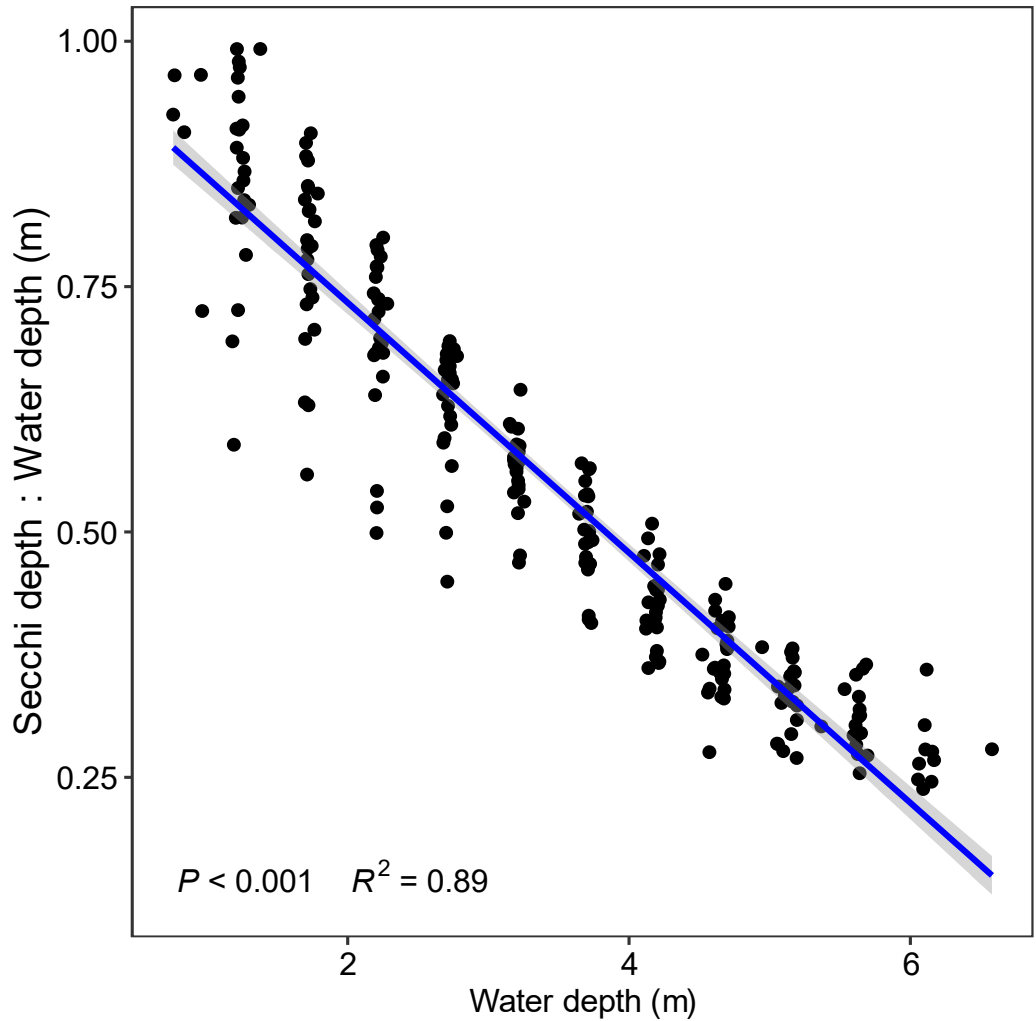

**Figure Supplementary 4**. The relationship between measured water depth using a sonic depth finder and Secchi disc measurements of depth (m). we used the ratio of water transparency to water depth to characterize underwater light conditions. Our results show that the ratio of water transparency to water depth decreases significantly with water depth, indicating that underwater light availability sharply decreases with water depth.

**Table Supplementary 1**. The name and relative abundance (RA) of dominant, common, and rare species in our study (19 species in total). Nomenclature follows the editorial committee of Chinese plant records.

| Dominant species | | Common species | | Rare species | |
| --- | --- | --- | --- | --- | --- |
| Latin name | RA | Latin name | RA | Latin name | RA |
| *Potamogeton maackianus* | 50.9 | *Hydrilla verticillata* | 5.2 | *Stuckenia pectinata* | 0.5 |
| *Ceratophyllum demersum* | 16.4 | *Potamogeton lucens* | 4.8 | *Potamogeton perfoliatus* | 0.4 |
| *Vallisneria natans* | 11.8 | *Myriophyllum spicatum* | 3.5 | *Potamogeton distinctus* | 0.3 |
| | | *Elodea canadensis* | 2.5 | *Potamogeton intortifolius* | 0.2 |
| | | *Potamogeton wrightii* | 2.2 | *Potamogeton acutifolius* | 0.2 |
| | | *Charophyceae* | 1.1 | *Najas marina* | 0.1 |
| | | | | *Utricularia aurea* | 0.01 |
| | | | | *Potamogeton crispus* | <0.01 |
| | | | | *Najas minor* | <0.01 |
| | | | | *Ottelia acuminata* | <0.01 |

Table Supplementary 2. Results of the Structural Equation Model testing the influence of water depth on potential stabilizing drivers of the temporal stability of the community biomass in the shallow water area. Given are the standardized path coefficients, standard errors of regression, critical value, and the levels of significance for the regression.

| Path (Shallow water area) | | | Standard coefficient | Standard error | Crit.Value | P.Value |
|---|---|---|---|---|---|---|
| Community stability | ← | Species richness | -0.05 | 0.01 | -0.83 | 0.41 |
| Community stability | ← | Species dominance | 0.22 | 0.17 | 4.17 | 0.00 |
| Community stability | ← | Species asynchrony | 0.41 | 0.18 | 7.33 | 0.00 |
| Community stability | ← | Dominant species stability | 0.85 | 0.06 | 13.66 | 0.00 |
| Community stability | ← | Common species stability | 0.40 | 0.05 | 8.23 | 0.00 |
| Community stability | ← | Rare species stability | 0.10 | 0.10 | 2.20 | 0.03 |
| Community stability | ← | Water depth | 0.00 | 0.02 | -0.02 | 0.98 |
| Species dominance | ← | Water depth | 0.12 | 0.01 | 1.47 | 0.14 |
| Dominant species stability | ← | Water depth | 0.36 | 0.03 | 5.79 | 0.00 |
| Common species stability | ← | Water depth | -0.25 | 0.03 | -3.33 | 0.00 |
| Rare species stability | ← | Water depth | -0.43 | 0.02 | -5.19 | 0.00 |
| Species asynchrony | ← | Water depth | 0.16 | 0.01 | 1.95 | 0.05 |
| Species richness | ← | Water depth | 0.15 | 0.18 | 2.58 | 0.01 |

**Table Supplementary 3**. Results of the Structural Equation Model testing the influence of water depth on potential stabilizing drivers of the temporal stability of the community biomass in deep water area. Give are the standardized path coefficients, standard errors of regression, critical value, and the levels of significance for the regression.

| Path (Deep water area) | | | Standard coefficient | Standard error | Crit.Value | P.Value |
|---|---|---|---|---|---|---|
| Community stability | ← | Species richness | 0.02 | 0.01 | 0.47 | 0.64 |
| Community stability | ← | Species dominance | 0.15 | 0.18 | 2.95 | 0.00 |
| Community stability | ← | Species asynchrony | 0.20 | 0.11 | 3.61 | 0.00 |
| Community stability | ← | Dominant species stability | 0.89 | 0.04 | 23.66 | 0.00 |
| Community stability | ← | Common species stability | 0.22 | 0.10 | 4.65 | 0.00 |
| Community stability | ← | Rare species stability | 0.05 | 0.13 | 1.87 | 0.07 |
| Community stability | ← | Water depth | -0.05 | 0.03 | -0.79 | 0.43 |
| Species dominance | ← | Water depth | 0.13 | 0.01 | 1.86 | 0.06 |
| Dominant species stability | ← | Water depth | -0.51 | 0.03 | -7.76 | 0.00 |
| Common species stability | ← | Water depth | -0.01 | 0.02 | -0.13 | 0.90 |
| Rare species stability | ← | Water depth | 0.05 | 0.01 | 0.38 | 0.71 |
| Species asynchrony | ← | Water depth | -0.50 | 0.02 | -7.28 | 0.00 |
| Species richness | ← | Water depth | -0.69 | 0.12 | -13.85 | 0.00 |